# Evolutionary Dynamics of Giant Viruses and their Virophages

Dominik Wodarz

Department of Ecology and Evolutionary Biology, 321 Steinhaus Hall, University of California, Irvine, CA 92697

Email: dwodarz@uci.edu; phone: 949-824-2531

## Abstract

Giant viruses contain large genomes, encode many proteins atypical for viruses, replicate in large viral factories, and tend to infect protists. The giant virus replication factories can in turn be infected by so called virophages, which are smaller viruses that negatively impact giant virus replication. An example are Mimiviruses that infect the protist Acanthamoeba and that are themselves infected by the virophage Sputnik. This paper examines the evolutionary dynamics of this system, using mathematical models. While the models suggest that the virophage population will evolve to increasing degrees of giant virus inhibition, it further suggests that this renders the virophage population prone to extinction due to dynamic instabilities over wide parameter ranges. Implications and conditions required to avoid extinction are discussed. Another interesting result is that virophage presence can fundamentally alter the evolutionary course of the giant virus. While the giant virus is predicted to evolve towards increasing its basic reproductive ratio in the absence of the virophage, the opposite is true its presence. Therefore, virophages can not only benefit the host population directly by inhibiting the giant viruses, but also indirectly by causing giant viruses to evolve towards weaker phenotypes. Experimental tests for this model are suggested.

**Introduction**

Mimivirus (microbe mimicking virus) was first discovered in the water of a cooling tower in the UK, infecting the protist *Acanthamoeba*, and was shown to have characteristics that are atypical for the majority of viruses (La Scola et al. 2003; Raoult et al. 2004; Koonin 2005; Claverie et al. 2006; Suzan-Monti, La Scola, and Raoult 2006; Raoult and Forterre 2008; Claverie and Abergel 2009, 2010; Colson and Raoult 2010; Forterre 2010; Yamada 2011). It was found to be a dsDNA virus and was the largest virus known at the time. Its 1.2 Mb genome sequence contained more than 900 proteins with functions that are not normally associated with viruses, such as encoding crucial components of the protein translation machinery (Raoult et al. 2004). Unlike other viruses, it was visible with a light microscope (Claverie and Abergel 2010; Sun et al. 2010). Mimivirus is thought to be phylogenetically close to other large DNA viruses (Claverie and Abergel 2009). They replicate in large viral factories that are reminiscent of simple cell nuclei, resulting in the lysis of their *Acanthamoeba* host. A different strain of Mimivirus with a slightly larger genome, called Mamavirus, was found in a different cooling water in France(La Scola et al. 2008). In this case, an interesting discovery was the association of Mamavirus with a small satellite virus that was named Sputnik (La Scola et al. 2008). Sputnik replicates within the viral factories of Mimiviruses, using Mimivirus resources and consequently impairing Mimivirus replication, leading to the generation of defective Mimivirus particles (La Scola et al. 2008; Pearson 2008; Claverie and Abergel 2009; Desnues and Raoult 2010; Ruiz-Saenz and Rodas 2010; Sun et al. 2010; Desnues et al. 2012; Zhang et al. 2012). This also reduces Mimivius-induced lysis of amoebae. Therefore, Sputnik is a true "parasite" of Mimivirus rather than a regular satellite virus and has consequently been termed a "virophage", although this distinction has been debated (Herrero-Uribe 2011; Krupovic and Cvirkaite-Krupovic 2011; Desnues and Raoult 2012; Fischer 2012).

Giant viruses and virophages are thought to be abundant in aquatic environments, infecting a variety of protists (Claverie et al. 2009; La Scola et al. 2010; Culley 2011; Yau et al. 2011). Consequently, virophages could play important roles regulating the population dynamics between protists and their viruses. This has been examined in Antarctic lakes, where a relative of the Sputnik virophage was found to infect phycodnaviruses, which in turn infect phototrophic algae (Yau et al. 2011). In this system, data analysis and population models suggested that virophages reduce the mortality of algal cells and that they could have an important influence on the stability of microbial food webs.

The impact of virophages on the dynamics between giant viruses and their host cells is related to the effects of hyperparasites on parasite-host dynamics. Hyperparasites are defined as parasites that infect another parasite, leading to a food chain of parasitism. The effect of hyperparasitism on population dynamics has been examined in some detail with mathematical models (Beddington and Hammond 1977; May and Hassell 1981; Hochberg, Hassell, and May 1990; Holt and Hochberg 1998), and the analysis often examined the impact on the biological control of insect pests. For example, Beddington and Hammond (Beddington and Hammond 1977) analyzed a scenario where a herbivore was infected by a parasite that was itself subject to infection by a hyperparasite. A recurrent result is that the introduction of a hyperparasite can reduce the effectiveness of biological control (Beddington and Hammond 1977; May and Hassell 1981). Because the primary parasite is attacked by the hyperparasite, the host/pest population benefits and can achieve higher equilibrium levels (Beddington and Hammond 1977; May and Hassell 1981). In addition, hyperparasites can influence the stability of a parasite-host system (Beddington and Hammond 1977). A detailed analysis of the stability of the food chain dynamics has been provided by Holt and Hochberg (Holt and Hochberg 1998), demonstrating both stabilizing and destabilizing effects. Related food web systems have been studied, including interactions among hosts, parasites, and predators, e.g. (Roy and Holt 2008).

Here, I build on these concepts and analyze mathematical models that describe the dynamics between a host protist, a virus infecting the protist, and a virophage infecting the virus. While the virophage is also a virus, for simplicity the term virus will be used to refer to the primary virus of the protist host, in order to distinguish it from the virophage. The model will be constructed with the *Acanthamoeba*-Mimivirus-Sputnik system in mind, although the model is quite general and also applicable to other systems. No population dynamic data exist so far to taylor the model to a specific system or to parameterize it. Instead, the general properties of the dynamics are investigated, in particular concentrating on the evolutionary dynamics of both the virus and the virophage. I will examine the evolution of "virophage pathogenicity", i.e. the degree to which the virophage inhibits replication of the primary virus. The model suggest that while selection favors a higher virophage pathogenicity, the emergence of more pathogenic virophages can also significantly destabilize the dynamics, rendering the system prone to extinction. Further, the evolution of the primary virus population is investigated. It is found that the evolutionary trajectory of the primary virus can be changed by the presence of the virophage. While in isolation, the primary virus is expected to evolve towards higher basic reproductive ratios, the presence of the virophage can lead to the evolution of the primary virus to a reduced basic reproductive ratio. Experiments with the *Acanthamoeba*-Mimivirus-Sputnik system are suggested to test and refine the model, as well as to estimate parameters.

**The mathematical models**

We consider an ordinary differential equation model that describes the average development of populations over time. These include the host Acanthamoeba population, *x*, amoebae infected with the Mimivirus, *y₁*, and amoebae infected with the Mimivirus which in turn is infected with the Sputnik virophage, *y₁₂*. Free virus is not explicitly taken into account. Since the life-span of viruses tends to be significantly shorter than that of cells, the virus populations are assumed to be in quasi steady state. The model is given by the following set of equations.

(1)
$$\frac{dx}{dt} = rx\left(1 - \frac{x + y_1 + y_{12}}{k}\right) - \beta_1 x(y_1 + fy_{12})$$
$$\frac{dy_1}{dt} = \beta_1 x(y_1 + fy_{12}) - a_1 y_1 - beta_2 \, y_1 y_{12}$$
$$\frac{dy_{12}}{dt} = \beta_2 \, y_1 y_{12} - a_{12} y_{12}$$

The amoeba population is characterized by logistic, density-dependent growth, described by the term $rx(1-(x+y_1+y_{12})/k)$. The intrinsic growth rate is given by $r$ and the total amoeba population (uninfected + infected individuals) cannot exceed the carrying capacity $k$. Contact between the primary virus and uninfected amoeba cells leads to infection with a rate $\beta_1$. The primary virus can be released from two sources. Obviously, one source are cells infected with the primary virus alone, $y_1$. An additional source are cells that contain both the primary virus and the virophage, although they are likely to release the primary virus at a reduced rate. This is expressed by the parameter $f$, which describes the degree of primary virus inhibition by the virophage (i.e. the virophage "pathogenicity") and can vary between zero and one. If $f=0$, the primary virus cannot replicate at all in the presence of the virophage. If $f=1$, the replication of the primary virus is not inhibited by the virophage. Amoeba infected with the primary virus only, $y_1$, die with a rate $a_1$ and become infected with virophage upon contact with a virophage-containing cell with a rate $\beta_2$. Amoebae infected with both the primary virus and the virophage, $y_{12}$, die with a rate $a_{12}$. This death rate is determined both by the virophage and the primary virus. We assume that the primary virus contributes less to cell death in the presence compared to the absence of the virophage, due to inhibition of viral replication (parameter $f$). In addition, the virophage itself can cause cell death with a rate $a_{ph}$. Thus, the overall death rate of this cell population is given by $a_{12} = a_{ph} + fa_1$. We do not track amoeba cells that are infected with the virophage only, since the virophage cannot replicate without the primary virus.

In order to address questions concerned with population extinction, we also consider a stochastic version of this model by applying the Gillespie algorithm to these ODEs.

**Basic properties**

The host amoeba population grows if r>0 and reaches carrying capacity *k* in the absence of infection. The primary virus grows if its basic reproductive ratio is greater than one. This is given by $R_0^{(1)}=\beta_1 k/a_1$. In this case, the system converges to the following equilibrium in the absence of the virophage:

$$x^{(0)} = \frac{a_1}{\beta_1}$$
$$y_1^{(0)} = \frac{r(\beta_1 k - a_1)}{\beta_1(r + \beta_1 k)}$$
$$y_{12}^{(0)} = 0$$

Note that the faster the replication rate of the primary virus, $\beta_1$, the lower the equilibrium number of infected cells. When a virophage is added to the system, it can establish an infection if its basic reproductive ratio is greater than one. It is given by $R_0^{(ph)}=\beta_2 y_1^{(0)}/a_{12}$. It is determined by the replication rate of the virophage and the death rate of infected cells, and also by the equilibrium number of cells infected by the primary virus in the absence of the virophage. As mentioned above, this is inversely proportional to the replication rate of the primary virus.

Therefore, if the replication rate of the primary virus lies above a threshold, then $R_0^{(ph)} < 1$, and the virophage fails to establish an infection. If the virophage does establish an infection, then the system converges to an equilibrium that is given by a very lengthy second degree polynomial and hence not written out here.

The dependence of the equilibrium population levels on the model parameters is largely intuitive. The host amoeba population is regulated by the primary virus, and the primary virus population is regulated by the virophage. Thus, a more effective virophage can down-regulate the primary virus population, and this can in turn increase the equilibrium levels of the host amoebae, as described in previous studies on hyperparasitism (Beddington and Hammond 1977; May and Hassell 1981). However, because virophage-infected cells can also transmit the primary virus to host amoeba, virophage infection kinetics can at the same time lead to a reduction of the amoeba population, giving rise to a tradeoff (Figure 1). This is seen in the dependence of the equilibrium amoeba host population size on the death rate of virophage-infected cells. The lower the death rate of the cells, the larger the amount of virus released from these cells (higher burst size). The amount of successful primary virus replication in virophage-infected cells is determined by the parameter $f$. If the value of $f$ is very low and close to zero, then primary virus replication is negligible in virophage infected cells. Hence, a higher virophage burst size due to a lower death rate of these infected cells impairs the primary virus, which in turn increases the equilibrium level of the host amoeba. Thus, a lower virophage-induced death rate of cells increases the host population (Figure 1). In contrast, when $f>>0$, then a significant amount of primary virus replication still occurs in virophage-infected cells, and we see a one humped relationship (Figure 1). For higher virophage-induced death rates of cells, $a_{ph}$, a reduction in $a_{ph}$ leads to larger host equilibrium levels as before. The inhibition of the primary virus, which benefits the host, is the dominant effect here. For lower levels of virophage-induced death of cells, however, the trend reverses and lower values of $a_{ph}$ lead to lower host amoeba equilibrium levels. Now the higher burst size of the primary virus, brought about by the reduced rate of virophage-induced cell death, is the dominant factor and negatively impacts the host population.

As discussed above, if the basic reproductive ratios of the primary virus and the virophage are greater than one, and if r>0, then the equilibrium describing the persistence of the two viruses and the host is stable. The equilibrium is approached by damped oscillations, with the damping time and the extent of the oscillations depending on the model parameters. Previous work on hyperparasitism has shown that the introduction of the hyperparasite can both have a stabilizing and a destabilizing effect on the dynamics. We examined how the degree of virophage-mediated primary virus inhibition (i.e. the "virophage pathogenicity") influences the approach to equilibrium (Figure 2). The most pronounced oscillations and the longest damping times are observed for maximal virophage pathogenicity, i.e. if the degree of primary virus inhibition is maximal such that *f=0* (Figure 2). Reducing the degree of virophage pathogenicity (increasing *f*) greatly stabilizes the dynamics, leading to significantly shorter damping times (Figure 2). Thus, higher degrees of virophage pathogenicity correlate with less stable dynamics.

If oscillatory dynamics occur, population extinction can be observed in a stochastic setting. This was shown by performing stochastic, Gillespie simulations of the ODEs (Figure 3). The parameters and cases considered are equivalent to those in Figure 2. The stochastic simulations were started at the integer population levels that are closest to the equilibrium numbers predicted by the ODEs, as this minimizes the extent of oscillations. Nevertheless, we observe quick population extinction for *f=0*, i.e. for maximally pathogenic virophages (Figure 3). Long-term persistence was observed for higher values of *f*.

**Evolution of virophage pathogenicity**

Here we examine the evolutionary dynamics of the virophage and concentrate in particular on the evolution of "virophage pathogenicity", defined by the parameter *f*, describing the degree to which the primary virus can replicate when infected with the virophage. We introduce a second virophage strain into the above model, which is now formulated as follows.

$$\frac{dx}{dt} = rx\left(1 - \frac{x + y_1 + y_{12} + z_{12}}{k}\right) - \beta_1 x\left(y_1 + f_1 y_{12} + f_2 z_{12}\right)$$

(2)
$$\frac{dy_1}{dt} = \beta_1 x\left(y_1 + f_1 y_{12} + f_2 z_{12}\right) - a_1 y_1 - \beta_2 y_1 y_{12} - \beta_2 y_1 z_{12}$$

$$\frac{dy_{12}}{dt} = \beta_2 y_1 y_{12} - a_{12} y_{12}$$

$$\frac{dz_{12}}{dt} = \beta_2 y_1 z_{12} - \alpha_{12} z_{12}$$

Cells containing the primary virus can become infected by two virophage strains, and the respective virophage-infected cells are denoted by $y_{12}$ and $z_{12}$. The two strains only differ in their pathogenicity, which is denoted by $f_1$ and $f_2$. The death rate of these infected cells is thus given by $a_{12} = a_{ph} + f_1 a_1$ and $\alpha_{12} = \alpha_{ph} + f_2 a_1$. The basic reproductive ratio of virophage strain 1 is given by $R^1_0{}^{(ph)} = \beta_2 y_1^{(0)}/a_{12}$ or $R^1_0{}^{(ph)} = \beta_2 y_1^{(0)}/(a_{ph} + f_1 a_1)$. The expressions for strain 2 is $R^2_0{}^{(ph)} = \beta_2 y_1^{(0)}/\alpha_{12}$ or $R^2_0{}^{(ph)} = \beta_2 y_1^{(0)}/(\alpha_{ph} + f_2 a_1)$. Because increased pathogenicity reduces the replication of the primary virus, it also increases the life-span of the infected cell. This in turn leads to a higher total viral output of the virophage and thus to a higher basic reproductive ratio. In this model, the virophage strain with the higher basic reproductive ratio wins the competition, as demonstrated in Figure 4. Hence, the virophage population is expected to evolve to maximum pathogenicity, i.e. to $f=0$.

As shown in the previous section, an increase in virophage pathogenicity can lead to more extensive population oscillations and longer damping times, with $f=0$ characterized by the most unstable dynamics. This can render populations prone to extinction, and these aspects were explored with stochastic Gillespie simulations of the ODEs. Figure 5 shows a scenario where a virophage strain with increased pathogenicity invades the population, and displaces the competing strain. The ensuing population oscillations quickly drive the virophage population extinct, and the primary virus can also be driven to extinction in this process. Thus, while selection favors a virophage strain with increased pathogenicity, the population can evolve to a state in which it is very prone to extinction.

Whether extinction occurs for maximally pathogenic virophages ($f=0$) depends on the model parameters and this is explored systematically in Figure 6. Obviously, whether extinction occurs or not can depend on the initial conditions, but is least likely if the simulation is started around the equilibrium values. Hence, starting from the equilibrium (at the nearest integer number, since the simulation is stochastic), the simulation was run for a defined period of time and it was recorded whether virophage extinction occurred during this time frame. This was done for different parameter combinations and the outcome is color-coded in Figure 6. Persistence requires that the equilibrium population levels are sufficiently high such that the oscillatory dynamics do not lead to extinction. In this respect, the equilibrium number of primary virus-infected cells, $y_1$, is of particular importance. If the virophage drives this population extinct, then it depletes its own targets for replication. High population levels of primary virus-infected cells, and thus persistence, is promoted by slow spread of the virophage, i.e. by a slow virophage replication rate, $\beta_2$, and a fast virophage-induced cell death, $a_{ph}$ (Figure 6). In addition, persistence is promoted by a fast growth rate of the host amoeba population, $r$ (Figure 6). The replication rate of the primary virus, $\beta_1$, and the rate of cell death induced by the primary virus, $a_1$, only have relatively small effects on the outcome (Figure 6). Because the virophage will likely evolve towards faster replication kinetics, this suggests that evolutionary trajectories will bring the system into a parameter regime that renders the populations prone to extinction, unless the host amoebae replicate sufficiently fast to avoid this.

**Evolution of the primary virus**

Here, the evolutionary dynamics of the primary virus are investigated, concentrating on the viral replication rate, $\beta_1$, and the rate of virus-induced cell killing, $a_1$. A model with two primary virus strains is considered that compete for the same host population. Cells infected with the second strain of the virus are denoted by equivalent capital letters, i.e. cells infected with the second strain of the primary virus only are denoted by $Y_1$, and cells that also contain the virophage by $Y_{12}$.

$$(3)\begin{aligned}\frac{dx}{dt} &= rx\left(1 - \frac{x + y_1 + y_{12} + Y_1 + Y_{12}}{k}\right) - \beta_1 x(y_1 + fy_{12}) - \gamma_1 x(Y_1 + fY_{12}) \\ \frac{dy_1}{dt} &= \beta_1 x(y_1 + fy_{12}) - a_1 y_1 - \beta_2 y_1 (y_{12} + Y_{12}) \\ \frac{dy_{12}}{dt} &= \beta_2 y_1 (y_{12} + Y_{12}) - a_{12} y_{12} \\ \frac{dY_1}{dt} &= \gamma_1 x(Y_1 + fY_{12}) - b_1 Y_1 - \beta_2 Y_1 (y_{12} + Y_{12}) \\ \frac{dY_{12}}{dt} &= \beta_2 Y_1 (y_{12} + Y_{12}) - b_{12} Y_{12}\end{aligned}$$

The infection rate of the second strain primary virus is given by $\gamma_1$, and the death rate of cells infected with the second strain primary virus is given by $b_1$ in the absence of the virophage and $b_{12}$ in the presence of the virophage (where $b_{12}=a_{ph}+fb_1$). The basic reproductive ratio of the first strain is the same as before, i.e. $R_0^{(y)}=\beta_1 k/a_1$, and that of the second strain is given by $R_0^{(Y)}=\gamma_1 k/b_1$.

In the absence of the virophage, the primary virus strain with the larger basic reproductive ratio wins the competition, and thus evolution will maximize the basic reproductive ratio (subject to constraints that are not included in this model).

The situation is more complex in the presence of the virophage. If a strain is characterized only by a higher replication rate ($\beta_1$ or $\gamma_1$) it always wins the competition. The most obvious reason is that a faster replication rate increases the basic replicative fitness of the virus. In addition, however, a faster replication rate of the primary virus indirectly conveys a benefit by weakening the virophage. As shown in equilibrium expression $y_1^{(0)}$, a faster replication rate of the primary virus reduces its equilibrium level in the absence of the virophage, and thus reduces the basic reproductive ratio of the virophage. In fact, weakening the virophage can be more important than increasing the basic replication kinetics of the primary virus. This is illustrated as follows. Assume that the second primary virus strain replicates faster ($\gamma_1 > \beta_1$) and that it is also characterized by a higher death rate of infected cells ($b_1>a_1$). Further assume that the increase in the death rate of infected cells is greater than the increase in the viral replication rate. In this case, the basic reproductive ratio of the second primary virus strain, $R_0^{(2)}$, is lower than that of the first strain, $R_0^{(1)}$; this also lowers the spread rate of the virophage. Under these assumptions, three outcomes are possible (Figure 7). As expected, the strain with the larger $R_0$ can win the competition. Interestingly, the strain with the smaller $R_0$ can also win and exclude its competitor. Alternatively, coexistence of the two strains can be observed. The dependence of the outcomes on parameters is explored in Figure 8a. Coexistence occurs only if $\gamma_1 \gg \beta_1$. If $R_0^{(2)}$ lies below a threshold, the second strain fails to invade and goes extinct. If $R_0^{(2)}$ is higher but still below the value of $R_0^{(1)}$, then the second strain can invade and exclude the first strain. The more effective the virophage is, the larger the parameter space in which the primary virus with the lower $R_0$ excludes the strain with the higher $R_0$. This is shown in Figure 8b-d by exploring the parameter space for different scenarios that vary in the effectiveness of the virophage. Therefore, if the virophage has a significant negative impact on the primary virus population, selection can favor primary viruses with a reduced $R_0$ because it lessens the impact of the virophage. In other words, the presence of the virophage can lead to evolution towards reduced replicative fitness of the primary virus.

**Discussion and Conclusion**

This paper used mathematical models to study the dynamics between a host population, its primary virus, and a virophage infecting the primary virus. In particular, the model was built with the Acanthamoeba-mimivirus-sputnik system in mind, although population dynamic measurements or parameter measurements that would allow a closer application are currently not available. Ecological studies point to the importance of virophages in regulating primary viruses and thus impacting protist populations (Yau et al. 2011). In the context of one specific study a Lotka-Volterra type mathematical model was used to underline this point (Yau et al. 2011), but this model was constructed to address specific questions about the ecology or arctic lake protist and was not meant to provide a more general model that systematically explores possible outcomes of the dynamics between hosts, primary viruses, and virophages. This was done here, with an emphasis on the evolutionary dynamics. Not surprisingly, some of the basic properties of the model are very similar to those observed in models of hyperparasitism (Beddington and Hammond 1977; May and Hassell 1981; Hochberg, Hassell, and May 1990; Holt and Hochberg 1998). For example, by regulating the primary virus population, the virophage can have a positive effect on the host amoeba population. However, since in the current model the virophage-infected cells can still allow transmission of the primary virus to host cells, a lower virophage-induced death rate of cells can also negatively impact the amoeba host population. A lower vriophage-induced death rate of cells not only allows release of more vriophages, but also of more primary virus.

Beyond the basic dynamics, some interesting evolutionary insights emerged. While the virophage is expected to evolve towards higher levels of primary virus inhibition, this can lead to more oscillatory dynamics that can lead to extinction of the virophage and also the primary virus. For pathogenic virophages, persistence is only possible for relatively slow virophage replication kinetics, which is again not favored by evolution. This brings up the question whether the presence of a virophage in food chains in transient and eventually destined to go extinct as a result of virophage evolution itself. This might not occur if there are constraints on evolution, i.e. if the virophage cannot evolve towards sufficiently high levels of pathogenicity and replication, due to factors not taken into account in the current model. Alternatively, it is possible that spatial interactions are required to ensure long-term virophage presence, because spatial structure tends to dampen the oscillatory dynamics in such systems, thus reducing the chance of extinction.

With respect to the evolution of the primary virus, the model suggests that the presence virophages can fundamentally alter the evolutionary course, selecting for primary viruses with a reduced basic reproductive ratio. This in turn could allow the ecosystem to evolve to a state that is beneficial for the host amoeba population.  Thus, the virophage may not  only benefit the host amoebae directly by attacking the primary virus, but it may also do so indirectly by influencing the course of primary virus evolution.

While our model was constructed specifically with virophages in mind, it could potentially also apply to satellite viruses in general.  There is a debate in the literature whether virophages represent a new class of viruses or whether they are part of the larger group of satellite viruses that require the help of another virus for replication (Herrero-Uribe 2011; Krupovic and Cvirkaite-Krupovic 2011; Desnues and Raoult 2012; Fischer 2012). It has been argued that some satellite viruses can also negatively impact the helper virus (Krupovic and Cvirkaite-Krupovic 2011). However, the model discussed here examines the role of virophage "pathogenicity" where the virophage can have a substantial impact on the fitness of the primary virus. Unless this assumption applies to satellite viruses, the applicability of the model presented here is limited.

In order to gain a more detailed understanding about the potential impact of virophages on the dynamics between the primary virus and its host, the model needs to be validated and experimentally tested. This could be done by in vitro experiments using the Acanthamoeba-mimivirus-sputnik system. First, it needs to be tested whether the model formulation presented here can accurately describe time series of Mimivirus infection in the absence and presence of the virophage. In other words, the validity of the equations used here needs to be established and possibly revised. Once the model has been validated, certain predictions about the evolutionary dynamics could be tested by generating different primary virus and virophage strains, and running competition experiments. Obtaining a detailed assessment of the population dynamic impact of virophages will be important for a better understanding of the microbial ecology of aquatic and marine systems.

**Figure Legends:**

**Figure 1:** Effect of virophage-induced cell death, $a_{ph}$, on the equilibrium host population, according to model (1). Different curves are shown, varying the virophage pathogenicity, $f$. Explanations are given in the text. Parameters were chosen as follows. $r=0.01$; $\beta_1=2.5 \times 10^{-7}$; $a_1=0.01$, $\beta_2=2 \times 10^{-5}$; $k=5 \times 10^5$.

**Figure 2.** Dynamics predicted by model (1), depending on the virophage pathogenicity, $f$. The host population is shown in black, the primary virus in blue, and the virophage in red. The more the virophage inhibits the primary virus (lower $f$), the more unstable the dynamics become, leading to more extensive oscillations and longer damping times. Parameters were chosen as follows. $r=0.01$; $\beta_1=2.5 \times 10^{-7}$; $a_1=0.01$, $\beta_2=2 \times 10^{-5}$; $a_{ph}=0.05$; $k=5 \times 10^5$.

**Figure 3.** Dynamics predicted by the stochastic, Gillespie simulation of ODE system (1), depending on the virophage pathogenicity, $f$. The host population is shown in black, the primary virus in blue, and the virophage in red. Figure 2 showed that dynamics become more unstable for lower $f$. Here, simulations were started at the equilibrium levels predicted by the ODEs (the nearest integer number) and typical outcomes were plotted. Starting around the equilibrium minimizes the chances of extinction due to oscillatory dynamics. For $f=0$, the dynamics are the most unstable and the system crashes to extinction. Higher values of $f$ stabilize the dynamics, resulting in long-term persistence. Parameters were chosen as follows. $r=0.01$; $\beta_1=2.5\times10^{-7}$; $a_1=0.01$, $\beta_2=2\times10^{-5}$; $a_{ph}=0.05$; $k=5\times10^5$.

**Figure 4.** Virophage competition, according to model (2). The red line depicts a virophage with a higher pathogenicity (lower $f$), while the blue line depicts the virophage with a lower pathogenicity. The virophage with the higher pathogenicity (lower $f$) wins the competition. Parameters were chosen as follows. $r=0.01$; $\beta_1=2.5\times10^{-7}$; $a_1=0.01$, $\beta_2=2\times10^{-5}$; $a_{ph}=0.05$; $f_1=0.05$; $f_2=0$; $k=5\times10^5$.

**Figure 5.** Evolution of the virophage to a higher degree of pathogenicity can lead to population extinction, according to Gillespie simulations of model (2). The simulation is started with the first virophage strain (blue) around equilibrium. The second virophage strain with increased pathogenicity (red) is subsequently introduced, invades, and excludes its competitor. Now the dynamics start to oscillate (due to the higher level of virophage pathogenicity), and the population crashes to extinction. Parameters were chosen as follows. $r=0.01$; $\beta_1=2.5 \times 10^{-7}$; $a_1=0.01$, $\beta_2=2 \times 10^{-5}$; $a_{ph}=0.1$; $f_1=0.05$; $f_2=0$; $k=5 \times 10^5$.

**Figure 6.** Extinction versus persistence of a virophage with maximal pathogenicity ($f=0$), in dependence of model parameters. The graphs are based on Gillespie simulations of model (2). Simulations were started at the equilibrium (nearest integer number) according to ODE model (1). The simulations were run until a time threshold of 50,000 time units, and it was recorded whether the populations were extinct (red) or persisted (blue). The parameters indicated in the plots were randomly varied 100,000 times. Note that the borders between extinction and persistence can be fuzzy due to randomness in the outcomes. The exact picture depends on the time threshold when the simulation is stopped. Obviously, any stochastic simulation will end in extinction if it is run for long enough, irrespective of the parameters. However, in the blue parameter region, persistence lasts for a significantly longer time than in the red region. Base parameters were chosen as follows. $r=0.01$; $\beta_1=2.5 \times 10^{-7}$; $a_1=0.01$, $\beta_2=2 \times 10^{-5}$; $a_{ph}=0.05$; $k=5 \times 10^5$.

**Figure 7.** Competition between two primary virus strains in the presence of the virophage, according to model (3). The first strain shown in red has a lower basic reproductive ratio, $R_0$, than the second strain shown in blue. As can be seen from the graphs, the strain with the lower basic reproductive ratio can win the competition, lose the competition, or coexistence can be observed, depending on the parameters. Parameters were chosen as follows. $r=0.01$; $\beta_1=2.5\times10^{-7}$; $a_1=0.01$, $\beta_2=2\times10^{-5}$; $a_{ph}=0.05$; $f=0.1$ $k=5\times10^5$; (a) $\gamma_1=2\times\beta_1$; $b_1=5\times a_1$; (b) $\gamma_1=2\times\beta_1$; $b_1=15\times a_1$; (a) $\gamma_1=10^{-6}$; $b_1=0.3$.

**Figure 8.** Outcome of competition between two primary virus strains in the presence of the virophage, depending on the parameter values, according to model (3). Strain 2 is assumed to have a lower basic reproductive ratio, $R_0$, than strain 1. Strain 2 with the lower $R_0$ wins in the red parameter region. Strain 1 wins in the blue parameter region. Coexistence is observed in the cyan parameter region. The grey lines indicate the ratio of $R_0$ for strain 2 over that of strain 1. The lower this ratio, the lower the relative $R_0$ of strain 2. The grey lines show that the ratio of $R_0^{(2)}/R_0^{(1)}$ per se does not determine the outcome of competition (on the lines, the ratio is identical). Different outcomes can be observed for the same ratio $R_0^{(2)}/R_0^{(1)}$. The different graphs show the parameter exploration for different parameter values. Panel (a) is the base scenario. Panel (b) assumes a stronger virophage due to a faster virophage replication rate. Because the virophage is stronger, the parameter region in which the primary virus with the lower $R_0$ wins is larger. Panel (d) also shows a stronger virophage, this time indirectly due to a faster replication rate of the host population, demonstrating a similar effect. Panel (c) is done for a relatively low virophage pathogenicity, i.e. a high value of $f$. This reduces the life-span of infected cells because of less inhibition of primary virus replication, thus lowering the virophage burst size. Consequently, the parameter region in which strain 2 with the lower $R_0$ wins is reduced. Base parameters were chosen as follows. (a) $r=0.01$; $\beta_1=2.5\times10^{-7}$; $a_1=0.01$, $\beta_2=2\times10^{-5}$; $a_{ph}=0.1$; $f=0.1$; $k=5\times10^5$. (b) Same is (a) except $\beta_2=2\times10^{-4}$. (c) Same as in (a) except $f=1$. (d) Same as in (a) except $r=0.1$.

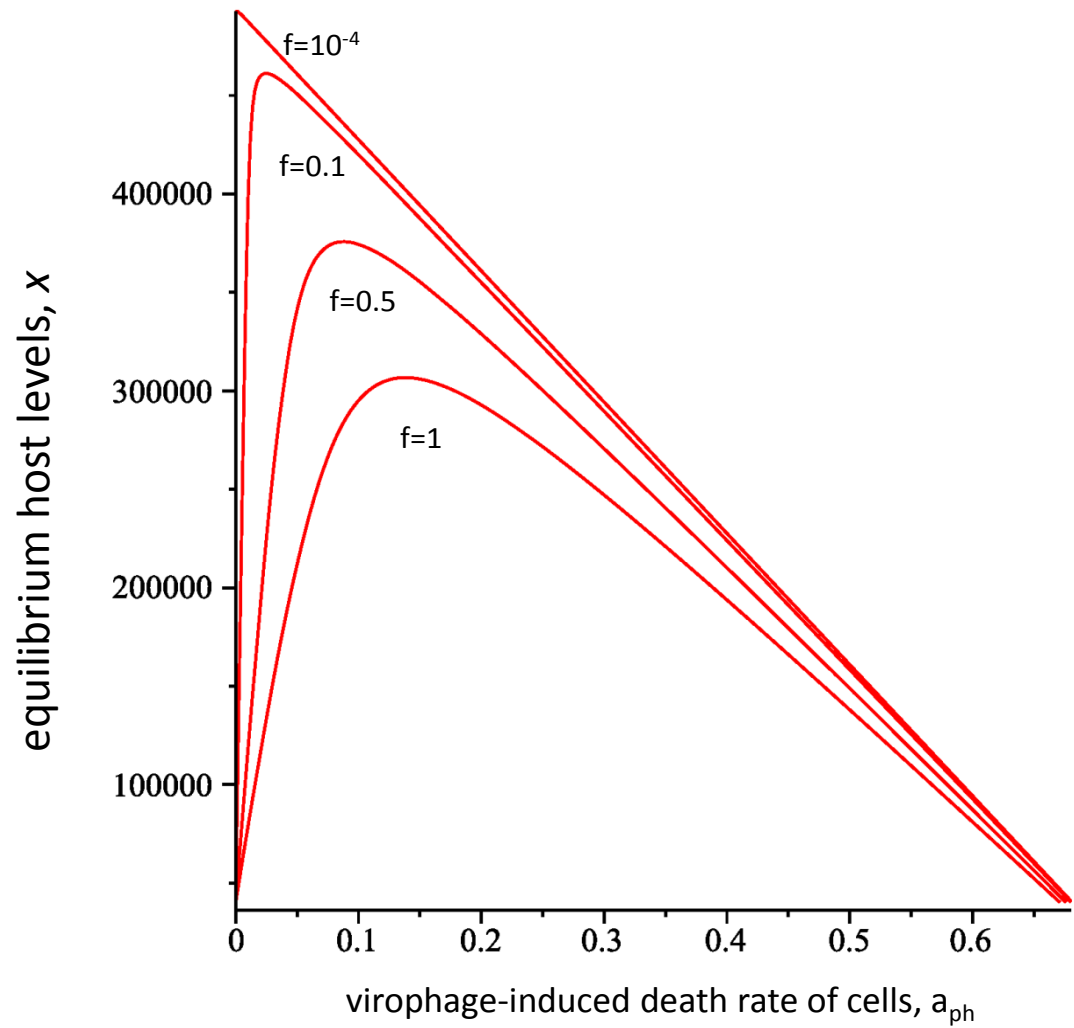

fig 1

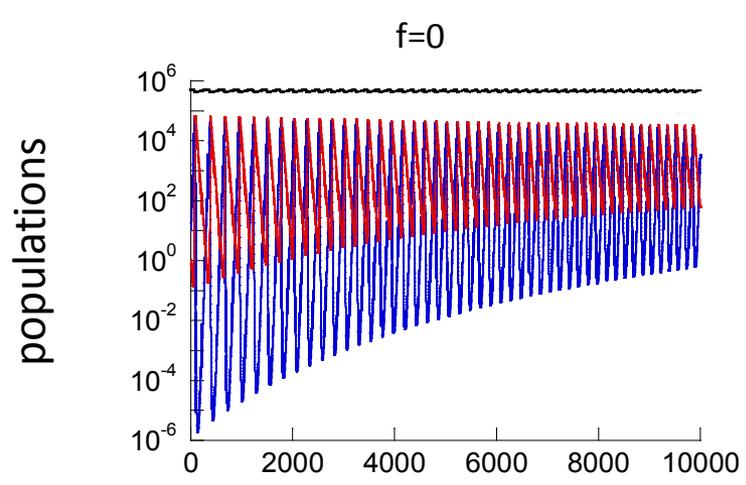
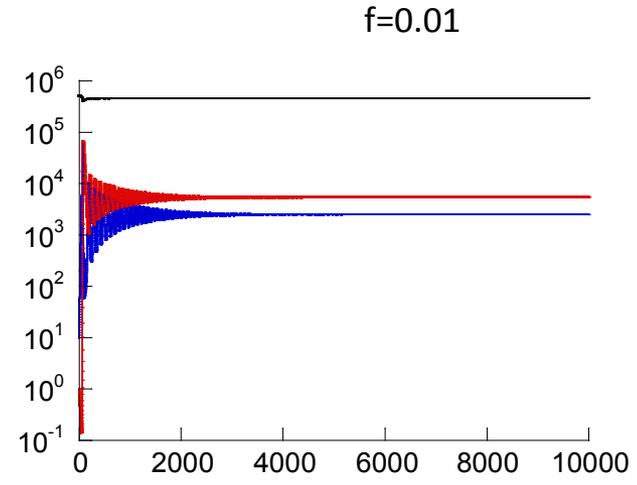
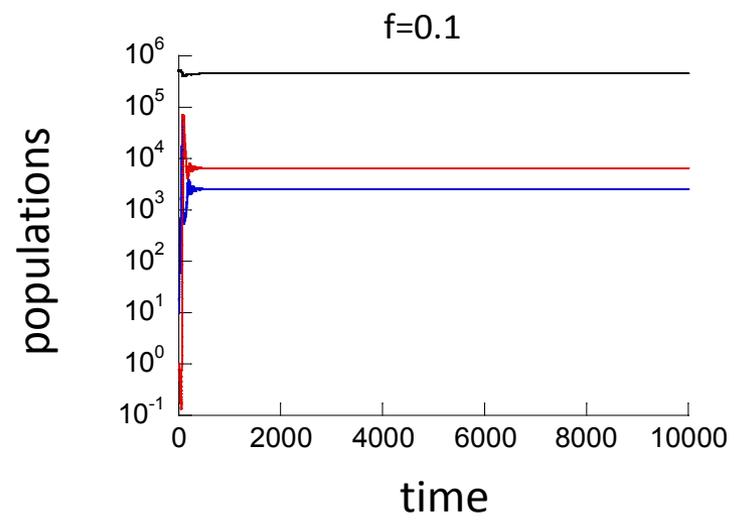
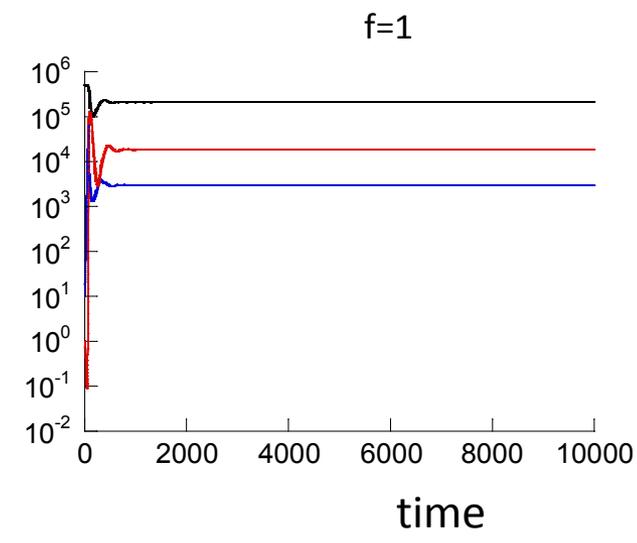

fig 2

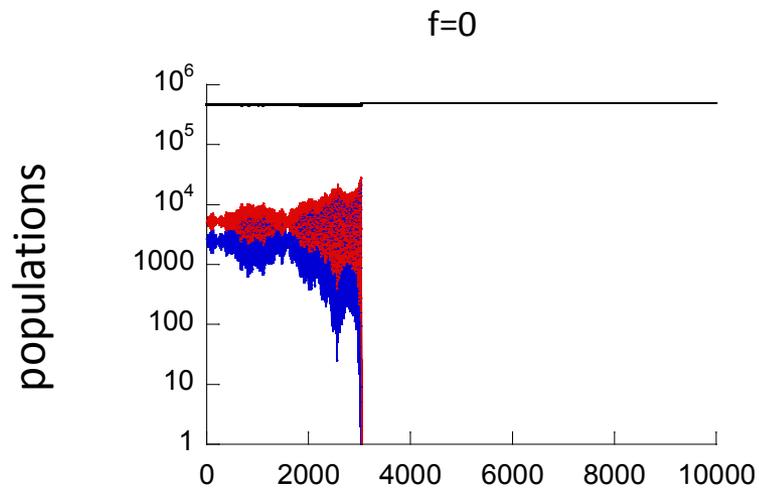
f=0

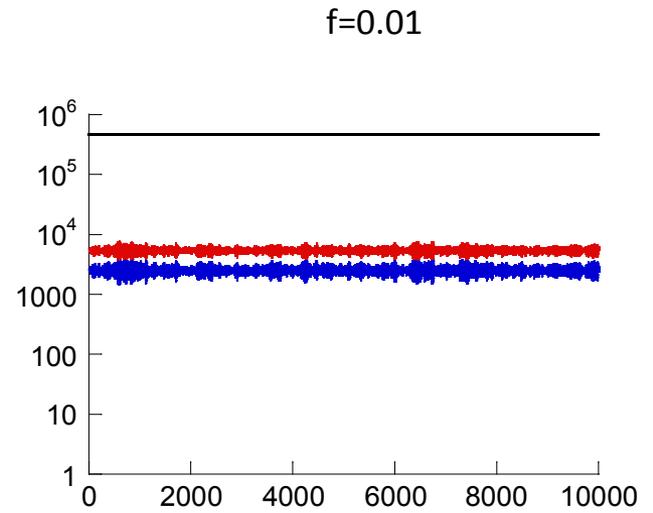
f=0.01

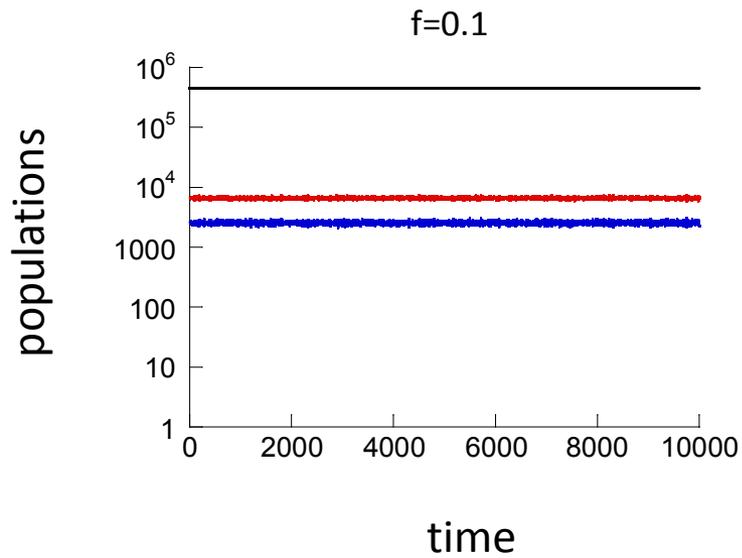
f=0.1

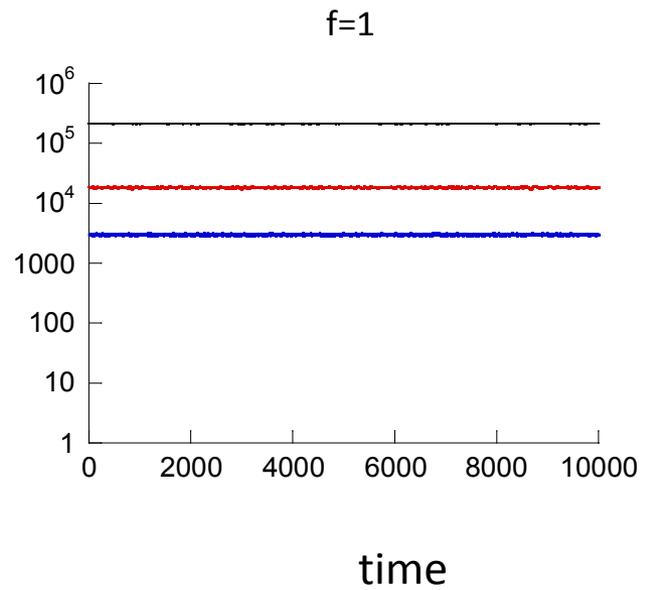
f=1

fig 3

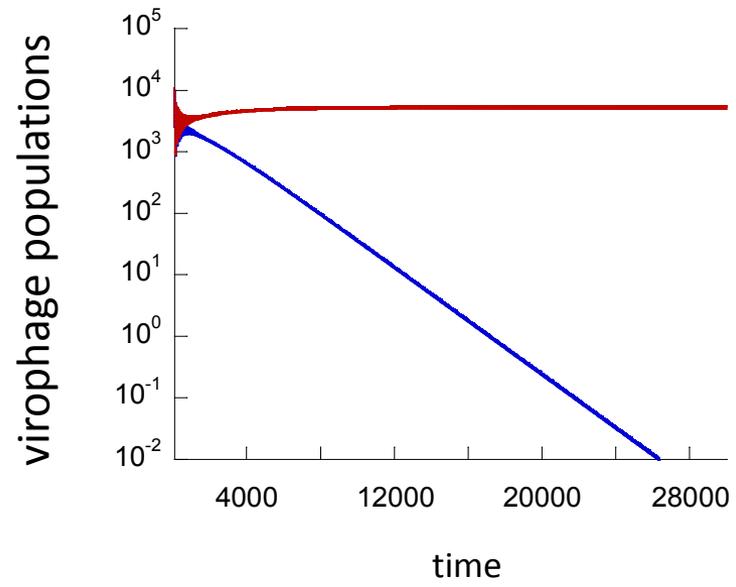

fig 4

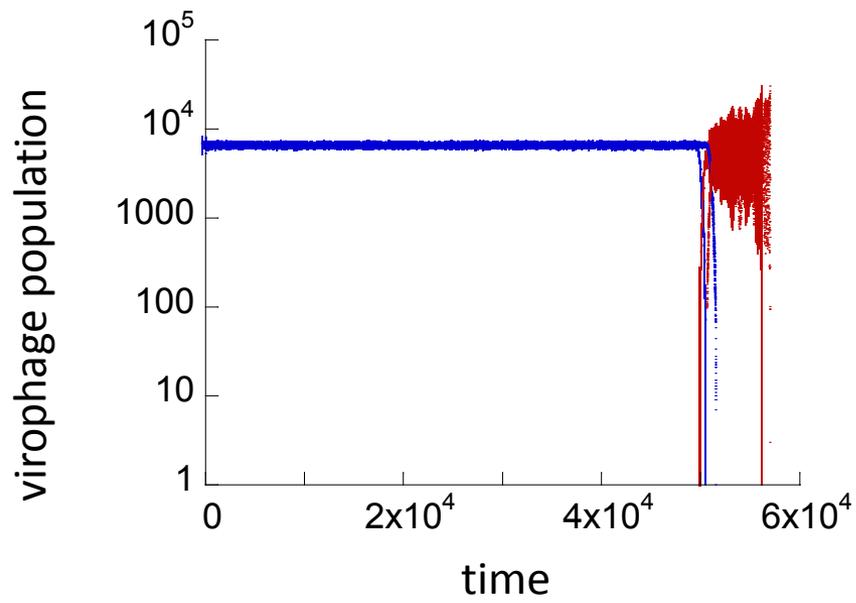

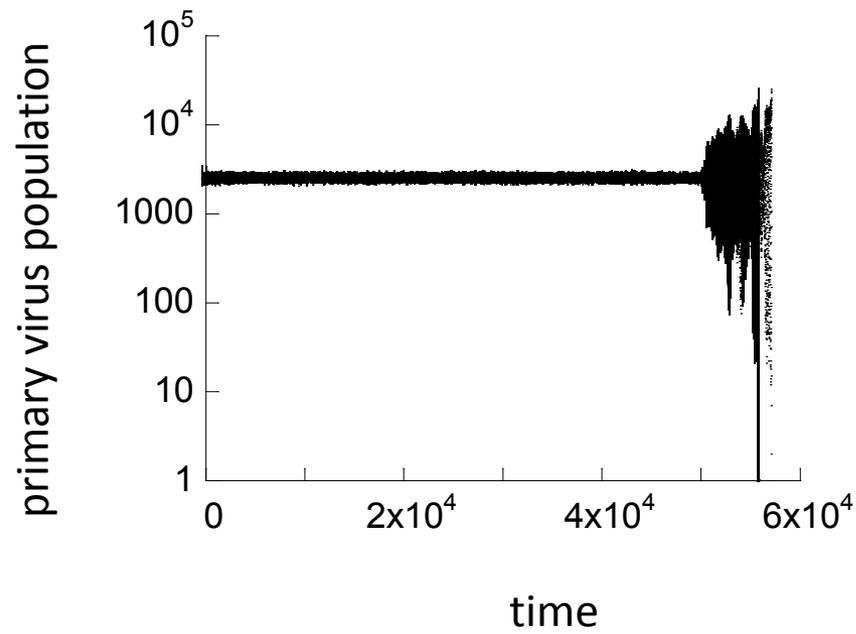

fig 5

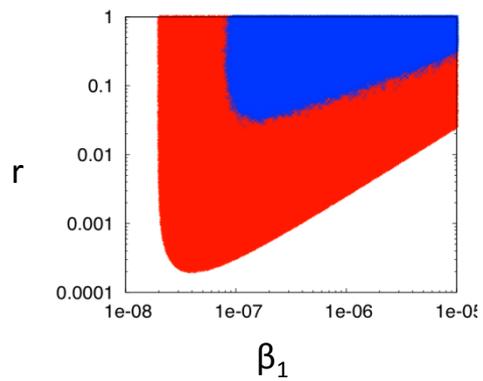
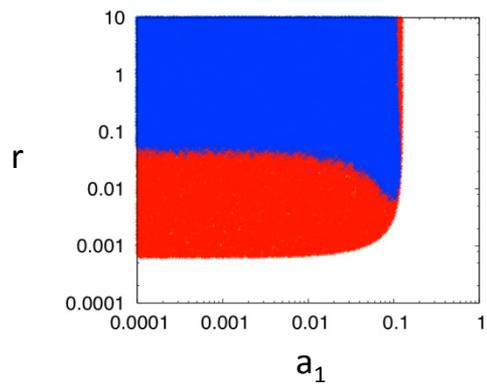
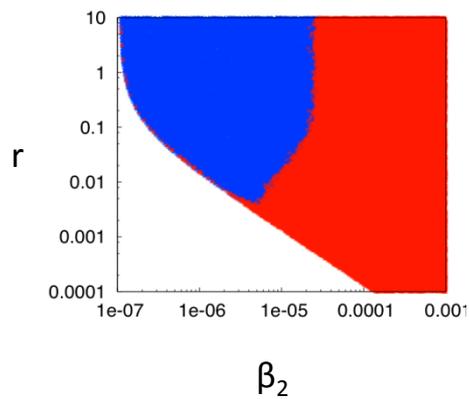
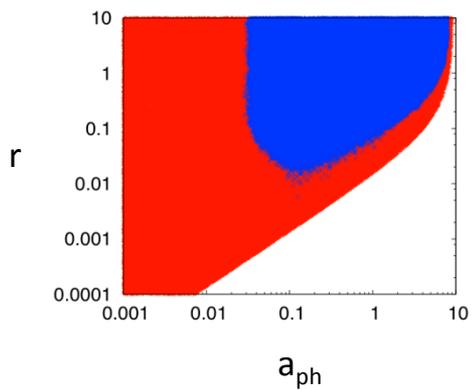
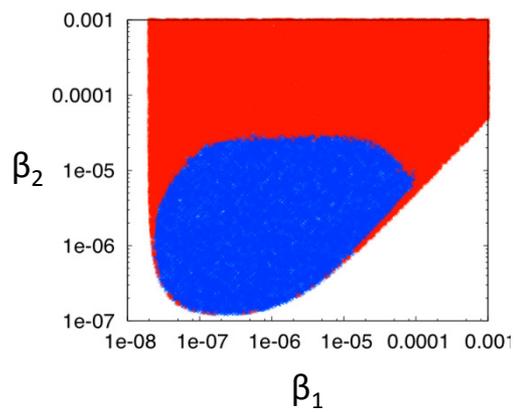

fig 6

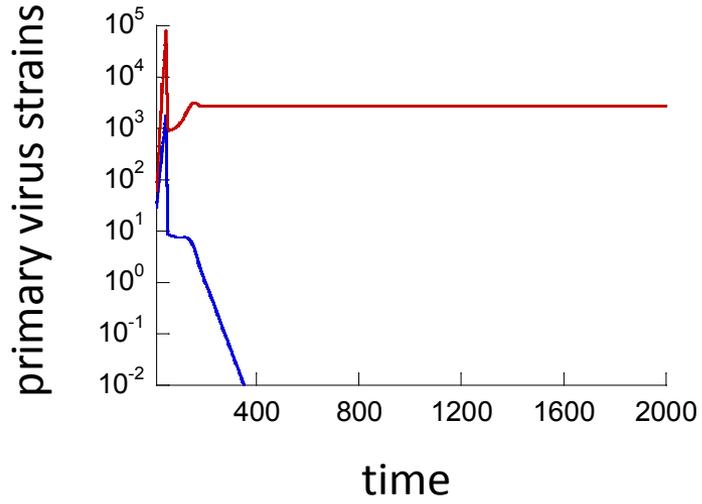
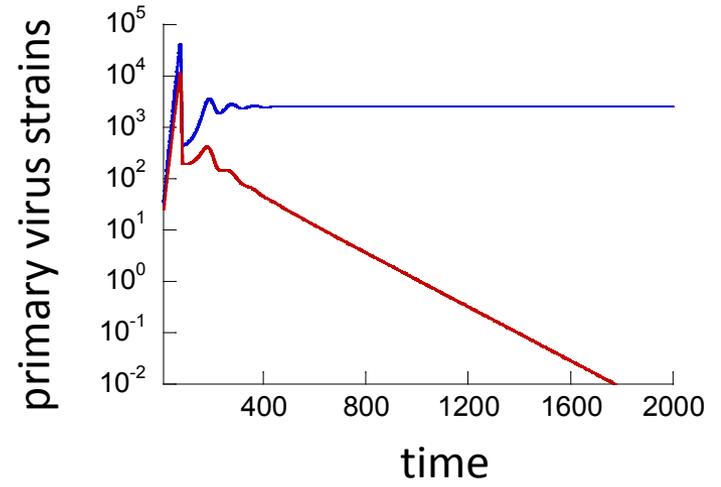
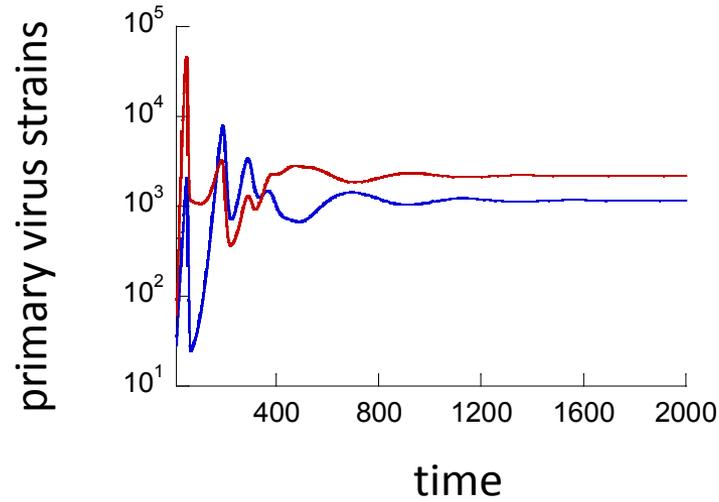

fig 7

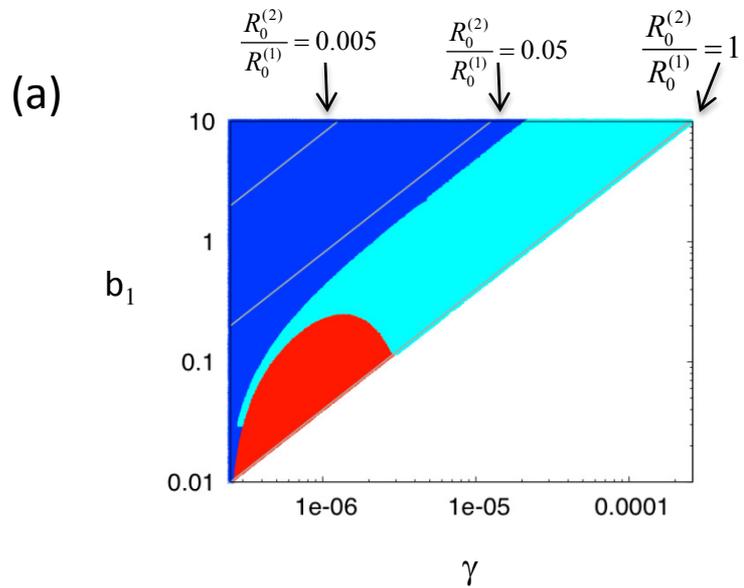
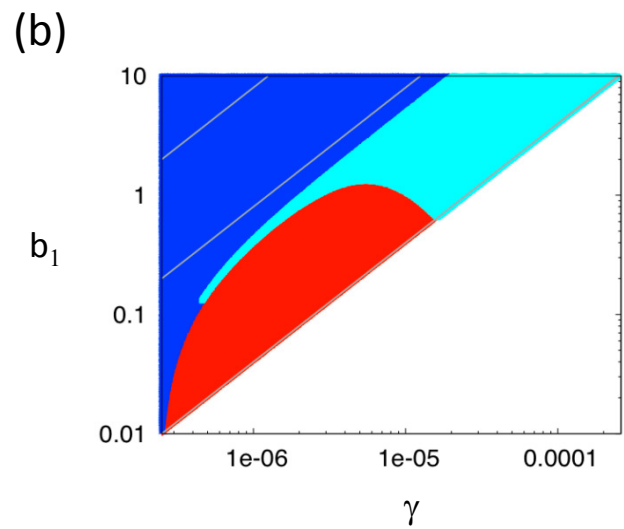
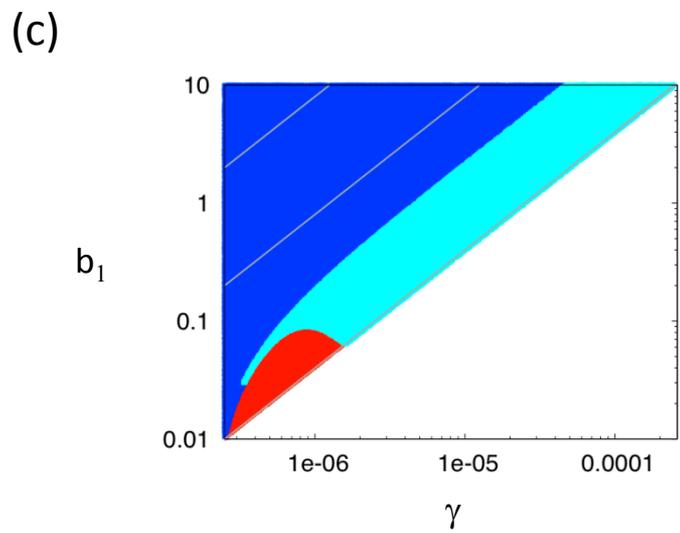
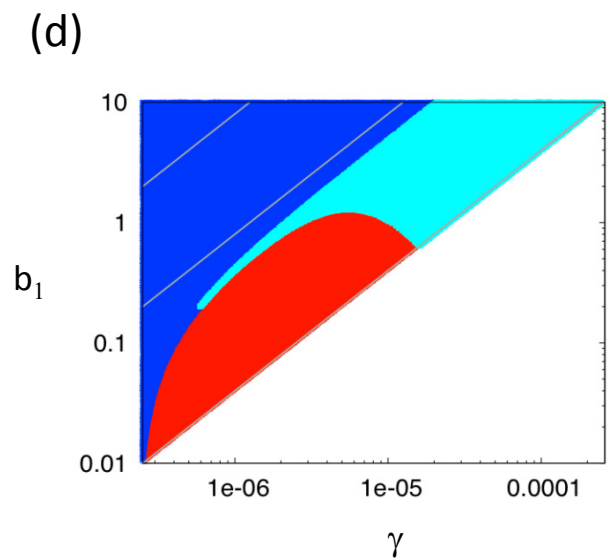

fig 8